# Early-stage local melting during solid-state sintering of a Cu-Ni-Ti-Zr-Hf alloy: A combined thermodynamic and kinetic analysis

Andrea Školáková[a,*], Hana Lesáková[a], František Lukáč[b], Markéta Jarošová[a], Jan Pinc[a]

[a] FZU – Institute of Physics of the Czech Academy of Sciences, Na Slovance 1999/2, Prague 8 182 00, Czech Republic

[b] Institute of Plasma Physics of the Czech Academy of Sciences, U Slovanky 2525/1a, 182 00, Prague, Czech Republic

*Corresponding author: skolakova@fzu.cz

## Abstract

Complex concentrated alloys (CCAs) combining elements from 3d transition and refractory groups, specifically Cu-Ni-Ti-Zr-Hf-based shape memory systems, exhibit excellent functional properties but present significant manufacturing challenges via powder metallurgy. In this study, a non-equimolar Cu15Ni35Ti25Zr12.5Hf12.5 (at. %) alloy was successfully produced by induction melting, followed by atomization and subsequent consolidation via Spark Plasma Sintering (SPS) at temperatures ranging from 850 to 1100 °C. Although equilibrium calculations using the current TCHEA7 database predict solid-state stability of the nominal composition up to slightly below 1100 °C, localized melting unexpectedly occurred during SPS already at 1000 °C and at 1100 °C. By combining retrospective Scheil-Gulliver solidification simulations with 2D local equilibrium heat maps, we demonstrate that this early-stage melting is not driven by macroscopic thermal gradients, but by microsegregation inherited from atomization. During solid-state heating, thermally activated short-range diffusion, accelerated by high defect densities at particle interfaces, partitions Ti towards the interparticle network, leaving Cu-enriched regions (up to ∼41 at. %) within the particle interiors. Owing to the strongly negative mixing enthalpy of Cu with Hf (-19.2 kJ/mol for $\Delta H_{mix}^{l-l}$) and, to a lesser extent, with Ti, the local solidus of these Cu-rich regions is dramatically depressed. Upon heating to 1100 °C, these regions cross their local solidus and form a liquid phase, whose solidification produces the Cu–Hf co-segregation. This framework accounts for localized melting phenomena in rapidly solidified CCA powders containing low-melting constituents with strong negative mixing enthalpies.

## Introduction

Conventional alloy development for advanced engineering applications has long been limited by traditional alloy design strategies, which typically rely on one primary solvent element with minor alloying additions. In contrast, complex concentrated alloys (CCAs) represent a shift in materials science by combining five or more principal elements in near-equimolar proportions [1]. Nowadays, rather than the traditional high-entropy alloy (HEA) concept, the more flexible complex concentrated alloy concept [1] is used, which groups all those alloys with a complex composition lying outside the traditional alloy design paradigm, irrespective of their microstructure. This multi-element approach introduces high configurational entropy, which favors stable solid-solution phases over complex intermetallic compounds. As a result, these alloys frequently exhibit high mechanical strength, thermal stability, and resistance to wear and corrosion. However, synthesizing such complex compositions through conventional casting often creates metallurgical challenges, including elemental segregation, dendritic microstructures, and casting defects. Powder metallurgy has emerged as an essential fabrication route to address these limitations. By providing strict microstructural control, uniform chemical composition, and near-net-shape manufacturing, powder metallurgy helps overcome processing barriers in complex concentrated alloys.

To date, research has focused primarily on Cantor-type alloys, typically based on 3d transition elements and their modifications [1-7], and refractory-element-based systems [1, 8-15]. A completely new class of complex concentrated alloys comprises those that combine elements from both categories with functional properties, specifically Cu-Ni-Ti-Zr-Hf-based shape memory alloys [16, 17]. Traditionally, CCAs have been produced exclusively via conventional melting routes [8, 18-20], however, powder metallurgy preparation has been increasingly explored in recent years. Conventional methods such as casting always lead to severe macrosegregation, the formation of undesired dendritic structures, or difficulties in combining elements with vastly differing melting points [21, 22]. Powder metallurgy overcomes these limitations [23]. CCAs are typically produced through a combination of mechanical alloying, in which a solid solution is formed by milling elemental powders, or by converting atomized as-cast alloys into powder, followed by compaction via Spark Plasma Sintering (SPS). This combination results in an ultrafine-grained microstructure while effectively suppressing grain growth and compositional segregation [18, 21, 24-26]. While the processing of 3d transition metal and refractory based alloys is well-established [18, 19, 21, 24, 25, 27-39], shape memory CCAs have not yet been successfully fabricated by this route. One of the major drawbacks is optimizing the processing parameters for both powder preparation and SPS compaction to achieve full densification, primarily because conventional thermodynamic data, such as those available for casting of CCAs, are lacking.

In this study, we investigated a Cu-Ni-Ti-Zr-Hf-based alloy. The material was first cast and subsequently atomized to avoid mechanical alloying, thereby saving tens of hours of preparation time, while maintaining the phase composition established during casting. Finally, the powder was consolidated via SPS. This work provides a combined experimental and theoretical explanation for early-stage local melting during SPS of powder mixtures at temperatures where equilibrium phase diagrams predict solid state stability. Given that prior thermodynamic prediction of such experimental behavior remains elusive and calorimeters capable of measuring above 1000 °C are not widely accessible, our study introduces a predictive framework based on retrospective Scheil solidification analysis. The One axis equilibrium phase diagrams traditionally favored in the literature are fundamentally unsuitable for these applications. Conversely, the Scheil model describes segregation in as-cast microstructures, a state that is eliminated during powder atomization, which means neither conventional tool can independently uncover the risk of melting. This risk is only revealed by combining the Scheil model for preferred redistribution pathways of the individual elements with elements heat maps. This approach applies to complex concentrated alloy and high entropy alloy powders containing an element with a low melting point and a strong negative enthalpy of mixing with another heavy primary constituent, such as Cu, Al, or Mg containing systems. It is similarly relevant to alloys with wide solidification ranges, in this case approximately 590 °C, and to rapidly solidified supersaturated powders consolidated near the local solidus temperature by methods including SPS, Hot Isostatic Pressing, Hot Pressing, or binder jetting sintering.

## Experimental methods

The non-equimolar Cu15Ni35Ti25Zr12.5Hf12.5 (at. %) alloy was prepared by induction melting of high-purity elemental components provided in cylindrical pellet form (ZhongNuo Advanced Material Technology Co., Ltd, China). The initial metal purities were specified as Cu 99.999%, Ni 99.995%, Ti 99.995%, Zr 99.95% (Zr + Hf ≥ 99.95%, Hf ≤ 0.03%), and Hf 99.95% (Zr + Hf ≥ 99.95%, Zr < 0.5%). The melting process was performed in an $Al_2O_3$ crucible under an argon atmosphere, with a total processing time of less than 5 min. This short processing time was selected to minimize the formation of secondary phases that could negatively affect the mechanical performance. The melt was subsequently poured into a steel mold with a diameter of 10 mm and a height of 120 mm. After casting, the ingot was left inside the furnace chamber to cool to ambient temperature. Subsequently, the bulk as-cast alloy underwent ultrasonic atomization using an ATO Lab+ US35 unit (3D LAB, Warsaw, Poland) equipped with a 35 kHz ultrasonic system under an inert argon environment. The feedstock was introduced via a Single Rod Feeding System (SRFS) and melted utilizing a TIG (tungsten inert gas) torch. The synthesized powder was subsequently classified with a Vibratory Sieve Shaker AS 200 control (Retsch® a VERDER company, Haan, Germany) to obtain the fraction < 63 µm. The particle size distribution of the atomized powders was measured using a SYNC particle analyzer (Microtrac Retsch GmbH, Germany), combining laser diffraction and dynamic image analysis. Wet (isopropanol as dispersion medium) measurement mode was employed. The distribution was characterized by d-values ($d_{10}$, $d_{50}$, $d_{90}$), defined as the particle diameters corresponding to 10%, 50%, and 90% of the cumulative mass fraction. Consolidation of the isolated powder was carried out via spark plasma sintering (SPS) in an SPS 10–4 system (Thermal Technology LLC, Nevada, USA) operating at 850 - 1100 °C, under an applied pressure of 100 MPa, with a heating rate of 500 °C/min (for the 1000 °C run, 500 °C/min up to 900 °C and 100 °C/min above). The pressure was applied before heating, and the temperature was measured and controlled by a thermocouple. To prevent carbon diffusion into the alloy, a tungsten foil barrier was placed between the samples and the graphite tools. This method was proven in [40].

Sample preparation involved mechanical grinding with SiC papers ranging from P180 to P4000. Final polishing was executed using a colloidal silica suspension (Eposil F) diluted with 30% $H_2O_2$ at a 1:6 volumetric proportion. The microstructure of the as-cast and SPS-compacted samples was investigated using an FEI 3D Quanta 3D field-emission gun DualBeam scanning electron microscope (SEM, FEI Company, Oregon, USA) equipped with a focused ion beam (FIB) for lift-out and an EBSD detector (TSL/EDAX Hikari). SEM imaging and EDS analyses of polished powder were performed on cross-sections using an Electron Probe microanalyzer (JEOL JXA-8230, JEOL Ltd., Tokyo, Japan) equipped with an energy-dispersive spectrometer (Bruker QUANTAX 200, Bruker Corporation, Massachusetts, USA). The chemical composition was determined at an accelerating voltage of 20 kV using a standardless P/B-ZAF quantification method. Data processing was carried out using ESPRIT 2.2 software. For each sample, at least six representative regions were analyzed by Energy-Dispersive X-ray Spectroscopy (EDS) to determine the chemical composition. The phase composition of the as-cast and powder materials was analyzed at ambient temperature using X-ray diffraction (XRD). Phase analysis was conducted using a Rigaku SmartLab 3 multipurpose diffractometer (Rigaku Corporation, Tokyo, Japan) and a PANalytical X'Pert PRO powder diffractometer (Cu anode, $K\alpha_1$ = 0.15406 nm, $K\alpha_2$ = 0.15444 nm; PANalytical, Almelo, Netherlands) operating in Bragg–Brentano geometry. The preliminary One axis calculation (**Fig. 3 a**), which was used to select the SPS temperature, was performed using Thermo-Calc software (version 2024b) with the TCHEA4 database for high entropy alloys. All other thermodynamic calculations, including the recalculated One axis diagram (**Fig. 3 b**), were performed using Thermo-Calc software (versions 2025a and 2025b) with the TCHEA7 database (v7.0) for high entropy alloys.

## Results and discussion

The microstructure of the as-cast alloy is shown in **Fig. 1 a**. As expected, it exhibits a dendritic structure, with Ni and Hf concentrated in the dendritic regions and Cu, Ti, and Zr in the interdendritic regions, as confirmed by the EDS elemental mapping. The chemical composition determined by EDS (**Fig. 1 c**) differed slightly from the nominal target for Cu (Cu was measured at 18 at. % instead of 15 at. %) and Ti (Ti was measured at 22.9 at. % instead of 25 at. %). This discrepancy can be attributed to these factors. The size of the interdendritic regions varied. Depending on the specific area analyzed during the EDS scan, a higher proportion of the Cu-rich interdendritic phase (**Fig. 1 c**) may have been included, leading to an artificially higher Cu concentration. For Ti, the characteristic Ti radiation was heavily absorbed by the dense matrix containing Hf and Zr, resulting in a slight underestimation of the Ti concentration, which can subsequently lead to incorrect recalculation by the software. Taking these factors into account, the actual chemical composition can be considered to match the nominal Cu15Ni35Ti25Zr12.5Hf12.5 (at. %) alloy within the experimental margin of error. In the as-cast state, the alloy was composed exclusively of martensite (B19') and austenite (B2), without any detectable secondary phases (see **Fig. S1** in Supplementary). The alloy was subsequently atomized, and the resulting powder was sieved to collect the < 63 µm fraction for microstructural analysis and subsequent compaction. Scanning Electron Microscopy - Backscattered Electron (SEM-BSE) imaging revealed no chemical contrast or elemental segregation, indicating a chemically homogeneous structure (**Fig. 1 b**). Because BSE imaging is directly sensitive to variations in average atomic number (*Z*), the absence of structural contrast confirms that the particles consist of a uniform, supersaturated solid solution without localized phase separation. This confirms that the high cooling rates during atomization induced solute trapping, freezing the alloy into a supersaturated solid solution. This was further supported by XRD analysis, which detected martensite and austenite phases, and an additional unidentified cubic phase (see **Fig. S2** in Supplementary). However, broad-area EDS chemical analysis revealed a notable shift in composition compared to the as-cast alloy, showing lower Cu and higher Ti contents (**Fig. 1 c**). The relative increase in Ti could also be largely a mathematical consequence of the apparent Cu depletion. The observed reduction in Cu concentration is theoretically attributed to compositional non-uniformity resulting from potential surface segregation toward the outer shells of the powder particles (see analyzed area in **Fig. 1b**). While quantitative EDS analysis near particle boundaries is inherently constrained by geometric edge effects, Cu enrichment at the shell remains a thermodynamically favorable scenario during atomization. Because Cu exhibits a substantially lower surface energy (1.83 J/m$^2$ [41, 42]) and melting point compared to Ni (2.45 J/m$^2$ [41, 42]), Ti (2.10 J/m$^2$ [41, 42]), Zr (2.00 J/m$^2$ [41, 42]), and Hf (2.15 J/m$^2$ [41, 42]), a strong thermodynamic driving force is expected to promote its migration to the surface to minimize total surface free energy. If present, these local compositional fluctuations would not yield detectable BSE contrast, as such chemical non-uniformity would form a nanoscale concentration gradient within a single solid-solution phase, lacking the physical thickness and Z-number variance required for electron backscatter imaging. Nevertheless, this metastable state could act as a potential ignition source for melting during subsequent SPS, which takes place essentially in the solid state, as was expected on the basis of the phase equilibria derived from the one-axis diagram. However, it should be added that when the atomized powder was analyzed by XRF (see **Table S1** in Supplementary), its overall composition was very close to the nominal alloy composition.

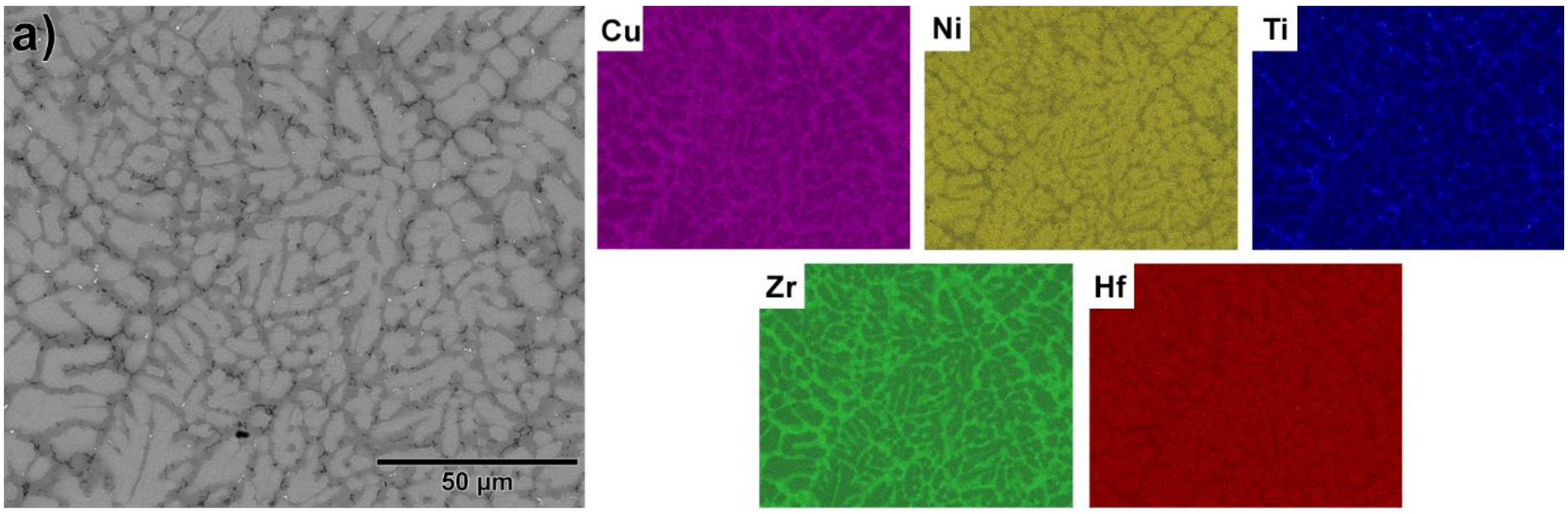

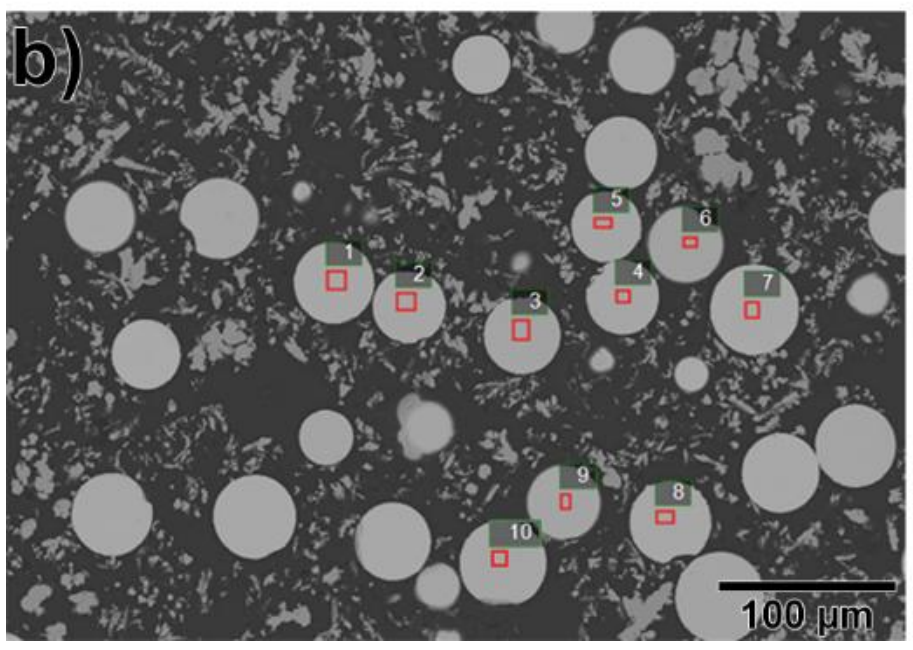

c)

| Element | Nominal composition | As-cast (overall) | As-cast dendrites | As-cast interdendritic | Atomized powder |
|---|---|---|---|---|---|
| Cu | 15.0 | 18.0 ± 2.8 | 14.8 ± 0.5 | 18.8 ± 2.9 | 9.2 ± 3.5 |
| Ni | 35.0 | 34.1 ± 0.5 | 34.6 ± 0.5 | 23.4 ± 4.6 | 35.0 ± 1.6 |
| Ti | 25.0 | 22.9 ± 0.4 | 20.4 ± 0.3 | 29.5 ± 7.3 | 30.4 ± 2.5 |
| Zr | 12.5 | 12.4 ± 1.1 | 9.0 ± 0.3 | 16.4 ± 2.4 | 12.9 ± 2.4 |
| Hf | 12.5 | 12.5 ± 1.8 | 21.2 ± 0.6 | 11.9 ± 3.0 | 12.4 ± 0.8 |

**Fig. 1:** a) BSE microstructure of the as-cast Cu15Ni35Ti25Zr12.5Hf12.5 (at. %) alloy alongside the corresponding EDS elemental distribution maps, b) BSE microstructure of the atomized powder, and c) measured chemical composition of the alloy and powder ("Overall" denotes an area analysis containing both dendritic and interdendritic areas)

The microstructure of Cu15Ni35Ti25Zr12.5Hf12.5 (at. %) alloy prepared under various SPS conditions is shown in **Figs. 2 a – e** with the corresponding chemical analysis results. None of the applied processing conditions achieved full consolidation of the material (the original spherical powder particles remained clearly distinguishable in the microstructure) in the solid state (without observed melting). Under the mildest condition (850 °C, 5 min) shown in **Fig. 2 a**, only partial necking of the initial atomized powder occurred. The particles largely retained their original spherical morphology, although localized regions exhibiting signs of consolidation lost their rounded shape. EDS analysis revealed significant compositional heterogeneity. While the overall bulk composition remained unchanged, elemental distribution maps showed that the residual spherical particles were markedly enriched in Cu, whereas the neck regions were enriched mainly in Ti and exhibited the lowest Cu concentration. The point analyses listed in the table (shown in **Fig. 2 a**) correspond to representative features (e.g., initial particles, interfaces, and region showing distinct BSE contrast). These point analyses confirm that most of regions of interest differ in chemical composition. Among the analyzed features, only the Cu concentration differs significantly, ranging from 4.9 to 14.6 at. % (**Fig. 2 a**). The concentration of Cu exhibited the most significant spatial variation across the microstructure, retaining its nominal alloy level exclusively within the initial particles, while being severely depleted in the surrounding regions. Increasing the temperature to 950 °C for 5 min (**Fig. 2 b**) still failed to obtain a fully consolidated structure. However, larger consolidated areas emerged, and some initial particles were fractured rather than remaining spherical. The consolidated regions showed Ti enrichment of more than 8 at. % relative to the nominal composition, alongside a lower Cu content. It could be assumed, that the possible surface layer enriched in Cu after atomization is consumed upon contact of particles, whereas the core retains its nominal Cu concentration, while Ti is present in the matrix. The Cu content in the cores reflects a relative depletion of the surrounding area rather than an absolute enrichment of the core itself. Bright consolidated regions observed in BSE mode were identified as Hf-rich zones. The concentrations of Ni and Zr remained relatively constant throughout these microstructural changes. Extending the holding time to 10 min at 950 °C broadened the inter-regional interfaces. These boundary regions were extremely enriched in Ti (reaching nearly 35 at. %), but their fine dimensions approach the interaction volume limit for accurate EDS analysis, making these measurements sensitive to chemical composition from the surrounding areas. The overall elemental distribution trends remained consistent with those observed at shorter holding times. The spherical particles retained their Cu enrichment with low Hf and Ti levels, whereas Ti was predominantly localized in the neck spaces. Increasing the SPS time to 20 min changed the microstructure substantially (**Fig. 2 d**), though full consolidation was still not achieved. The initial/original atomized powders were still observed, but were more fragmented. While Cu remained concentrated within these particles, its distribution became more uniform than in the previous state. However, across the samples sintered at 950 °C, EDS analysis even revealed regions containing up to 41 at. % Cu. According to elemental mapping, darker regions within the consolidated matrix were enriched in Ti and Zr, with lower contents of Ni, Hf, and Cu. Conversely, the brighter structural regions were depleted in Ti and enriched in Hf. The kinetic feasibility of this solid-state partitioning during sintering at 950 °C for 20 min can be verified by estimating the characteristic diffusion distance ($x = \sqrt{D \cdot t}$). Based on experimentally measured tracer diffusion in HfTiZrNbTa and HfTiZrNbV refractory high-entropy alloys [43-45], representative bulk diffusivities at 950 °C are expected to be on the order of D $\sim 10^{-15} - 10^{-14}$ $m^2/s$. A dwell of 20 min at 950 °C therefore corresponds to a lattice diffusion length of 1 – 3.5 µm. This is comparable to the thickness of the Ti- and Hf-enriched interparticle area but is an order of magnitude smaller than the particle radius (the mean particle size was: $d_{10}$ = 37.71 µm, $d_{50}$ = 50.89 µm, $d_{90}$ = 61.86 µm). Lattice diffusion alone can therefore account only for redistribution within a shell of a few micrometres around the particle contacts, not for transport across the particle interiors. Accordingly, the compositional heterogeneity observed within the particles is interpreted as inherited from atomization and subsequently reinforced near the contact zones, where grain-boundary and pipe diffusion, promoted by the high defect density introduced during compaction, are substantially faster than lattice diffusion. Overall, the microstructural evolution across all SPS conditions confirms persistent chemical inhomogeneities. The systematic variation of SPS parameters revealed that Cu, Ti, and most probably Hf played a key role in elemental redistribution, which ultimately led to localized melting as the temperature increased to 1000 °C and 1100 °C. It must be emphasized here that sintering at 950 °C proceeded in the solid state. In contrast, SPS at 1000 °C for 5 min already resulted in melting. The extruded melt caused failure of the die, so no sample could be recovered for microstructural analysis. Melting also occurred at 1100 °C. The SPS process records (Fig. S3 in Supplementary) document these melting events. At 1000 °C, an abrupt piston displacement of ∼0.8 mm occurred at ∼995 °C, i.e., before the set temperature was reached, accompanied by a temporary temperature stall despite increasing current. This indicates an endothermic melting event. At 1100 °C, an abrupt displacement of ∼1.7 mm occurred at ∼1030 °C during heating. In contrast, no such event was recorded at 950 °C, even after 20 min. No significant temperature overshoot was detected in any run (maximum 1003 °C and 1105 °C, respectively). The microstructure resulting from localized melting during SPS at 1100 °C is shown in **Fig. 2 e**. This structure closely resembles the as-cast state, featuring distinct dendritic and interdendritic regions.

However, in contrast to the as-cast microstructure, where Cu segregated to interdendritic regions, the dendrites formed during SPS consisted of Ni, Hf, and newly incorporated Cu. Because the potential surface oxide films, which could hinder the atomic transport during SPS, were not observed and EDS analysis did not find the presence of oxygen, the incomplete consolidation across these conditions can be attributed to the limited solid-state diffusion distance under the given time-temperature regime (see the estimate above), rather than to an intrinsically sluggish diffusion. The localized melting observed at 1100 °C was likely triggered by localized Cu enrichment, which decreased the local solidus temperature.

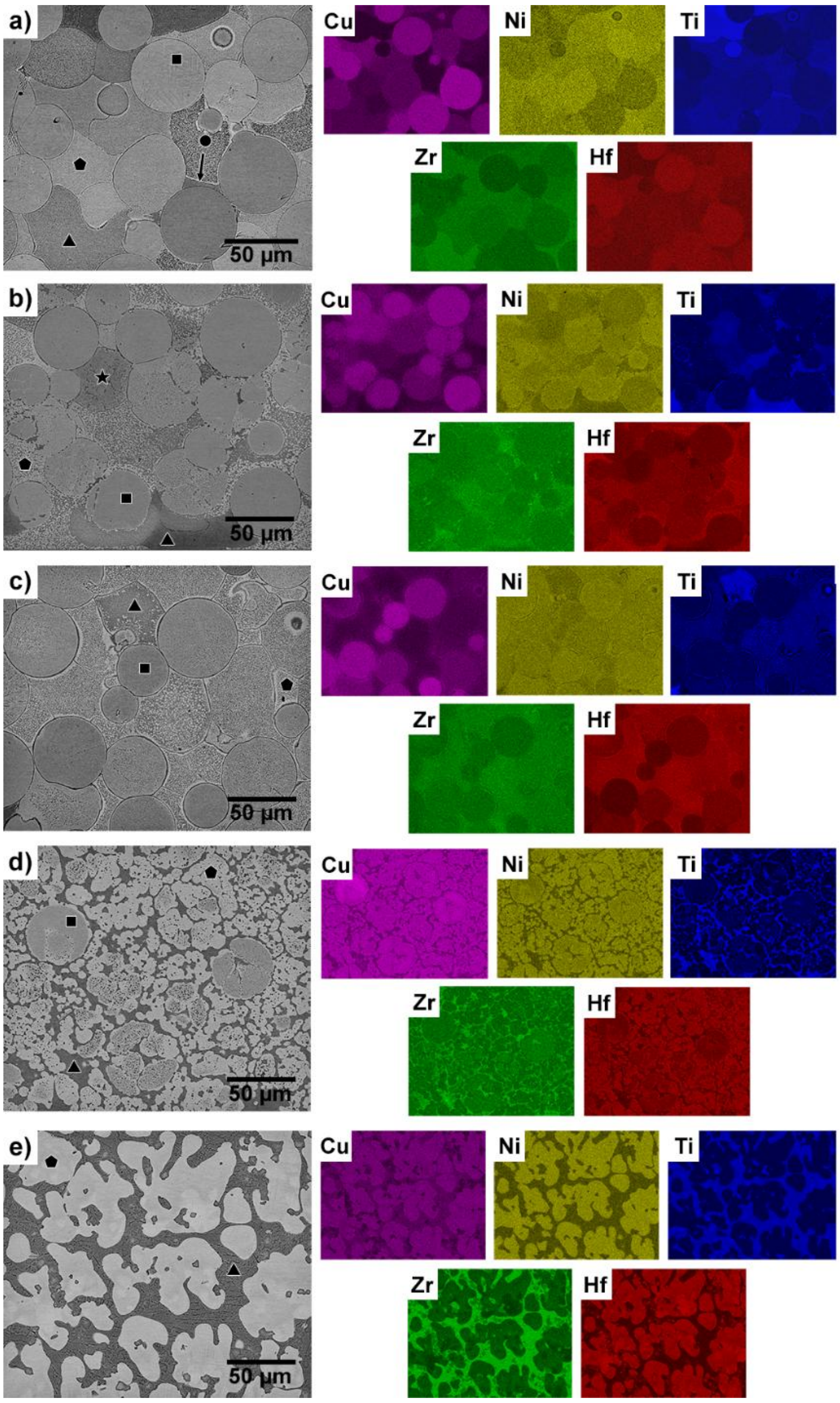

a)

| Element | ⬟ | ▲ | ■ | ● |
|---|---|---|---|---|
| | At.% | | | |
| Zr | 14.9 ± 1.7 | 14.5 ± 0.8 | 12.3 ± 0.7 | 11.1 ± 1.9 |
| Ti | 26.2 ± 1.7 | 26.2 ± 1.5 | 23.6 ± 0.2 | 23.4 ± 5.3 |
| Ni | 37.4 ± 1.4 | 34.8 ± 0.4 | 35.0 ± 0.5 | 38.4 ± 2.2 |
| Hf | 16.5 ± 1.1 | 13.6 ± 0.1 | 14.5 ± 1.3 | 14.3 ± 3.0 |
| Cu | 4.9 ± 1.5 | 10.8 ± 2.8 | 14.6 ± 1.4 | 12.7 ± 10.1 |

b)

| Element | ⬟ | ▲ | ■ | ★ |
|---|---|---|---|---|
| | At.% | | | |
| Zr | 11.0 ± 1.1 | 12.0 ± 0.5 | 12.0 ± 0.7 | 10.1 |
| Ti | 20.5 ± 4.0 | 33.9 ± 1.7 | 21.5 ± 0.6 | 26.2 |
| Ni | 36.1 ± 3.0 | 28.7 ± 1.1 | 32.7 ± 1.7 | 29.7 |
| Hf | 22.9 ± 3.6 | 17.4 ± 0.7 | 18.8 ± 1.0 | 16.8 |
| Cu | 9.5 ± 1.5 | 8.0 ± 1.2 | 14.9 ± 1.5 | 17.2 |

c)

| Element | ⬟ | ▲ | ■ |
|---|---|---|---|
| | At.% | | |
| Zr | 13.0 ± 1.2 | 12.8 | 11.6 ± 0.3 |
| Ti | 22.2 ± 3.3 | 30.1 | 21.9 ± 0.1 |
| Ni | 34.1 ± 1.3 | 30.9 | 32.8 ± 1.4 |
| Hf | 21.5 ± 2.5 | 20.4 | 18.6 ± 1.5 |
| Cu | 9.1 ± 2.6 | 5.8 | 15.0 ± 2.9 |

d)

| Element | ⬟ | ▲ | ■ |
|---|---|---|---|
| | At.% | | |
| Zr | 10.6 ± 0.9 | 14.1 ± 1.3 | 11.1 ± 1.0 |
| Ti | 18.3 ± 0.4 | 33.7 ± 0.5 | 21.3 ± 0.7 |
| Ni | 36.0 ± 0.4 | 27.8 ± 0.4 | 32.9 ± 1.1 |
| Hf | 21.7 ± 0.5 | 16.2 ± 1.2 | 18.7 ± 1.3 |
| Cu | 13.3 ± 0.9 | 8.2 ± 0.5 | 16.1 ± 3.4 |

e)

| Element | ⬟ | ▲ |
|---|---|---|
| | At.% | |
| Zr | 13.0 ± 1.7 | 13.2 ± 0.2 |
| Ti | 18.0 ± 1.8 | 37.9 ± 0.2 |
| Ni | 32.8 ± 3.3 | 25.3 ± 0.2 |
| Hf | 21.1 ± 1.5 | 16.9 ± 0.4 |
| Cu | 15.0 ± 2.6 | 6.7 ± 0.1 |

**Fig. 2:** SEM-BSE microstructure, EDS elemental distribution maps, and point analysis of the Cu15Ni35Ti25Zr12.5Hf12.5 (at. %) alloy prepared by SPS at a) 850 °C, 5 min, b) 950 °C, 5 min, c) 950 °C, 10 min, d) 950 °C, 20 min, e) 1100 °C, 5 min

The microstructure formed by localized melting at 1100 °C (**Fig. 2 e**) is dendritic and therefore resembles the as-cast state, but it forms from a melt whose composition is set locally rather than by the nominal alloy composition. The dendritic regions contain 21.1 at. % Hf and 15.0 at. % Cu, and the interdendritic regions 37.9 at. % Ti (**Fig. 2 e**). Because the volume that melted is a small fraction of the compact, its composition is not limited to the nominal one, and the partitioning observed here reflects solidification of a locally Cu- and Hf-enriched liquid. The origin of this local enrichment is analyzed in the following section.

**Equilibrium conditions**

To determine the cause of the unexpected melting below the nominal solidus temperature, the equilibrium phase evolution was analyzed and compared with the non-equilibrium processing history of the alloy. The calculated one-axis property diagram of the studied Cu15Ni35Ti25Zr12.5Hf12.5 (at. %) alloy (**Figs. 3 a, b**) illustrates theoretical phase evolution and temperature-dependent phase volume fractions assuming full thermodynamic equilibrium. According to the preliminary calculation using TCHEA4 database (**Fig. 3 a**), on the basis of which the SPS temperatures were selected, the liquidus temperature marking the onset of crystallization from the melt is located at approximately 1640 °C, whereas the solidus temperature reaches roughly 1260 °C. The current TCHEA7 database (**Fig. 3 b**) predicts a similar liquidus of approximately 1690 °C, but a considerably lower solidus, slightly below 1100 °C (only ~0.1 % of liquid is predicted at 1100 °C). This difference of more than 150 °C between the two database versions illustrates the sensitivity of the predicted solidus for the class of alloys. According to TCHEA7, the solidification temperature range is remarkably wide, nearly 590 °C. Conventionally, such a broad solidification range implies a strong tendency to extreme dendritic segregation during casting. Based purely on elemental melting points, high melting point elements would be expected to solidify preferentially within the dendrite cores, while lower melting point elements would partition into the interdendritic regions. However, the experimental elemental distribution map of the as-cast alloy contradicts this simplistic assumption (**Fig. 1 a**). Instead, the dendrites are predominantly enriched in Ni (melting point 1455 °C) and Hf (2233 °C), along with some Ti (1668 °C), whereas the interdendritic regions accumulate Cu (1085 °C), Ti, and notably Zr (**Fig. 1 a**), despite Zr possessing a high melting point of 1855 °C. This discrepancy demonstrates that elemental redistribution during rapid non-equilibrium casting is not determined by elemental melting points, but is governed primarily by the thermodynamic partition coefficients ($k$) and the mutual chemical affinities. Atomization of the dendritic as-cast alloy suppresses macro-scale segregation via rapid solidification. While SEM-BSE observation initially suggested a microstructurally homogeneous powder owing to the uniform atomic-number contrast (**Fig. 1 b**), point-by-point EDS analysis revealed underlying chemical fluctuations among individual particles (**Fig. 1 c**). Below approximately 950 – 970 °C, both calculations predict a miscibility gap within the B2 phase, where the matrix decomposes into two structurally identical yet chemically distinct phases designated as BCC B2 and BCC B2#2. A single-phase B2 region is predicted between approximately 970 and 1250 °C by TCHEA4 and between approximately 950 °C and slightly below 1100 °C by TCHEA7, theoretically suggesting the temperature interval for potential consolidation via SPS. However, experimental observations during powder compaction clearly reveal that melting occurs already at 1000 °C, i.e., below the solidus predicted by both database versions, and also at 1100 °C (**Fig. 2 e**). This breakdown of equilibrium predictions highlights the dominant role of non-equilibrium kinetics during powder processing, the underlying mechanisms of which are analyzed in detail in the following sections.

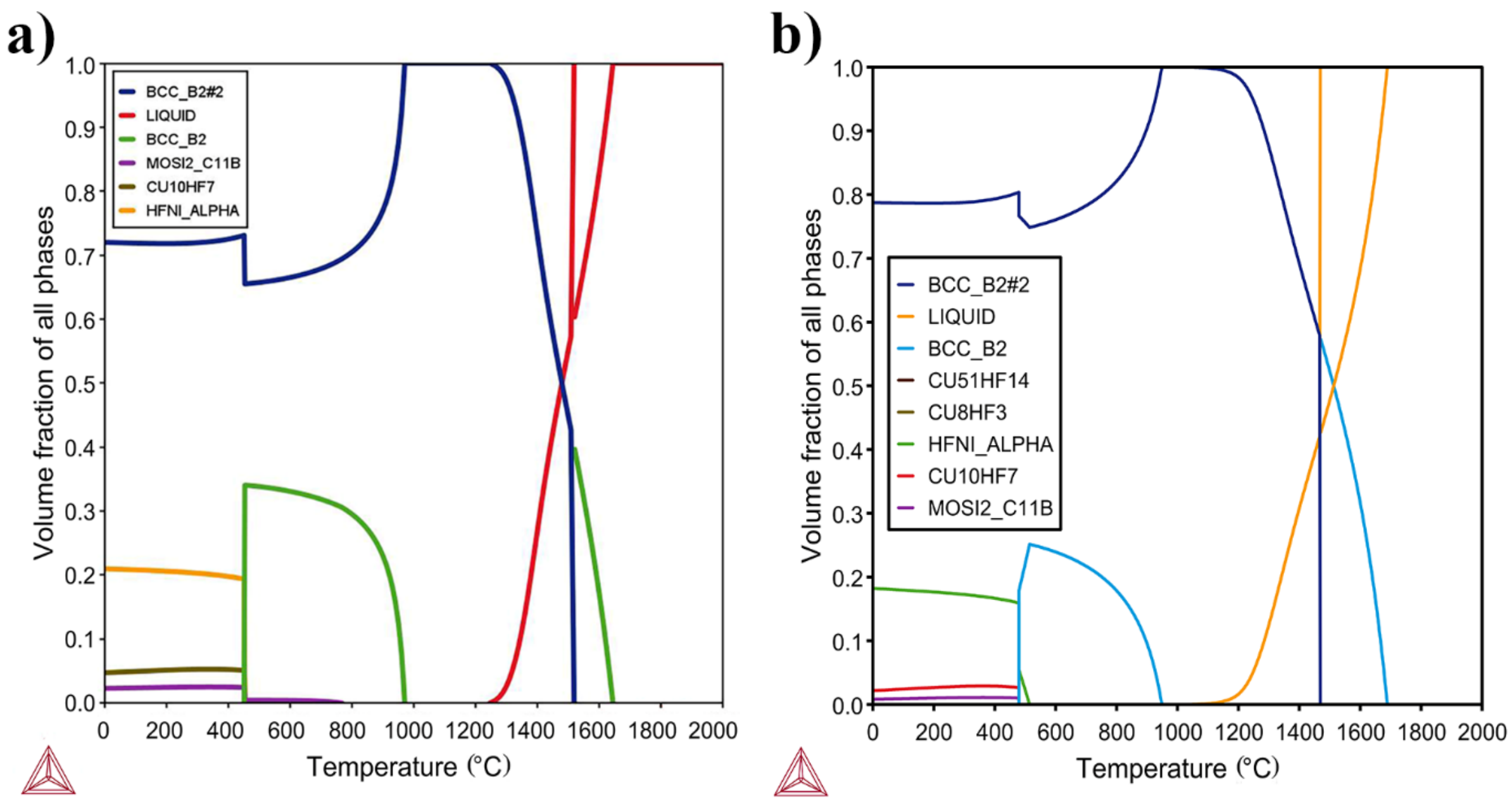

**Fig. 3:** One axis diagrams of the Cu15Ni35Ti25Zr12.5Hf12.5 (at. %) alloy calculated using a) TCHEA4 (preliminary calculations used for the selection of the SPS temperatures) and b) TCHEA7. In b), BCC B2 and BCC B2#2 above ~950 °C denote composition sets of the same B2 phase; the change of the label at ~1465 °C has no physical meaning

**Non-equilibrium conditions**

Since the real casting process occurs under non-equilibrium conditions, the solidification path and localized chemical trends were further evaluated using the Scheil–Gulliver model (**Fig. 4 a**). The classical Scheil equation describes solute redistribution during non-equilibrium solidification. In the present work, solidification was simulated using a CALPHAD-based Scheil–Gulliver approach, where local thermodynamic equilibrium and phase partitioning are recalculated at each solidification step using thermodynamic databases. This model assumes complete mixing in the liquid, negligible diffusion in the solid, and local equilibrium at the solid-liquid interface [46]. Under these assumptions, the solute distribution is governed by the partition coefficient,

$$k = \frac{c_S}{c_L},$$

while the solute concentration in the liquid during solidification follows

$$c_L = C_0(1 - f_S)^{k - 1},$$

where $c_0$ is the initial alloy composition, $f_s$ is the solid fraction, and $k$ is the equilibrium partition coefficient. Thermodynamic simulation of non-equilibrium solidification using the Scheil-Gulliver model revealed extreme microsegregation tendencies in the studied Cu15Ni35Ti25Zr12.5Hf12.5 (at. %) alloy and the explanation can be seen in **Fig. 4 a**. As solidification progresses, solute enrichment of the remaining liquid increases, resulting in pronounced microsegregation. **Fig. 4 a** contained equilibrium (dotted line) and non-equilibrium (colored line) solidification. Under equilibrium solidification conditions, the onset of crystallization occurs just below 1700 °C, while solidification is completed slightly below 1100 °C, in agreement with One axis calculation using the same database (**Fig. 3 b**). The steep final part of the equilibrium curve between approximately 1200 °C and 1100 °C corresponds to the solidification of the last small fraction of Cu-enriched liquid at $f_s \rightarrow 1$. According to the Scheil solidification model (colored line), the crystallization path can be effectively tracked, as the rapid cooling rate during casting inherently limits solid-state diffusion. According to this simulation path, BCC_B2 phase is the first to crystallize, growing over a wide temperature range down to 890 °C. Subsequently, a secondary phase designated as the $MoSi_2$_C11b structure type begins to crystallize together with BCC_B2 phase. This is driven by the heavy elemental enrichment of the residual liquid (green part of the line in **Fig. 4 b**). At approximately 860 °C, the curve was nearly horizontal, suggesting a reaction similar to a eutectic/peritectic transformation in binary systems. The remaining liquid phase solidifies and the crystallization of the Cu51Hf14 phase (**Fig. 4 a**) and the solidification was completed. Based on these results, the elemental distribution across the dendrites can be reliably predicted. As illustrated by the experimental results in **Fig. 1 a**, BCC_B2 phase preferentially depletes Ni and Hf from the melt due to their high mutual chemical affinity, forming primary dendrites enriched in both elements. With decreasing temperature, the dendrites coarsen and the solid fraction increases strongly. However, due to the rapid crystallization inherent to casting conditions, solid-state back-diffusion within the already existing dendrite cores is kinetically restricted. Consequently, the surrounding melt becomes severely enriched in the remaining solutes (Cu, Ti, and Zr), making the subsequent precipitation of the $MoSi_2$_C11b and Cu51Hf14 type phases in the interdendritic regions highly probable. However, diffraction did not reveal them due to their volume fraction being below the detection limit (~1 – 3%). Because Zr and Ti are chemically related and form isostructural phases [47], it can be assumed that these elements occupy the tetragonal lattice of the C11b type phase. Furthermore, Zr and Hf belong to the same group of the periodic table that share nearly identical atomic radii and chemical behavior. Although the Cu51Hf14 phase is conventionally defined as a binary intermetallic within the Cu - Hf system, extensive mutual substitution between Zr and Hf occurs within this complex concentrated alloy. The actual chemical composition of this phase is therefore more accurately represented as Cu51(Hf, Zr)14. XRD analysis of the as-cast alloy (see **Fig. S1** in Supplementary material) demonstrated that the near ambient temperature microstructure is predominantly composed of a monoclinic phase B19' (martensite) alongside a minor volume fraction of a cubic phase (retained austenite). The equilibrium calculation (**Fig. 3**) predicts a BCC_B2-type matrix as the dominant phase at low temperature, which is consistent with the parent phase identified by XRD. The monoclinic B19' martensite detected experimentally is the product of a diffusionless transformation of this parent phase and is not represented in the thermodynamic database. The comparison is therefore limited to the parent structure. The high concentrations of Ni and Hf within the dendrite cores effectively elevated the transformation temperatures above ambient temperature, resulting in a nearly complete transformation of the dendritic features. The detected retained austenite is localized in the transition zones at the dendrite peripheries, where the changed chemical composition induced by Scheil segregation locally stabilized the cubic BCC_B2 lattice below ambient temperature. The minor phases predicted at low temperature (HFNI_ALPHA, CU10HF7) were not detected by XRD, which is attributed to the equilibrium calculation not being representative of the rapidly solidified as-cast state.

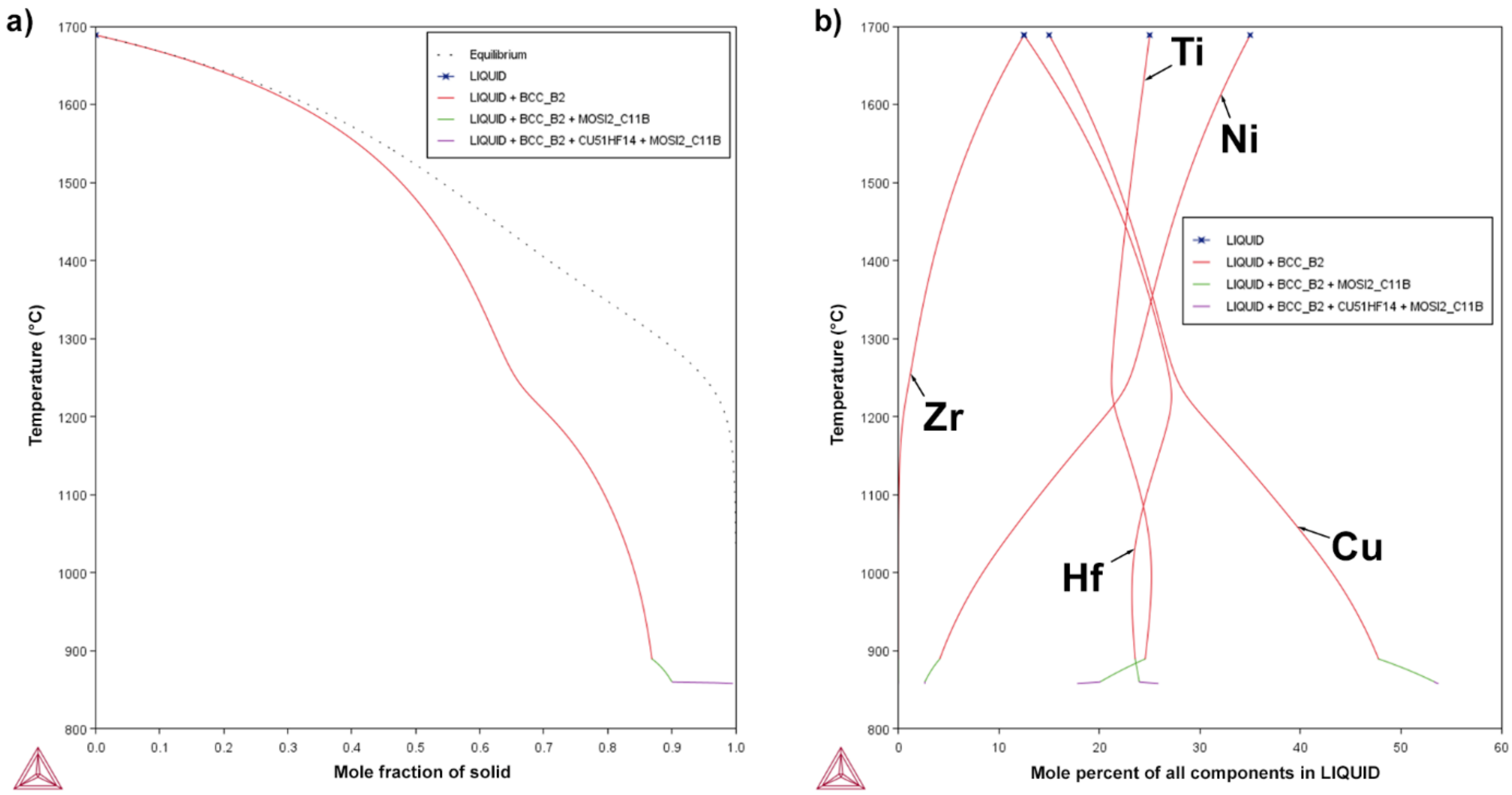

**Fig. 4:** a) Scheil-Gulliver solidification curve, b) temperature dependence of solute concentration in the liquid phase for Cu15Ni35Ti25Zr12.5Hf12.5 (at. %) alloy

The experimental elemental distribution observed in the as-cast state can be fundamentally understood by analyzing the constituent partitioning within the ordered BCC_B2 phase, which follows a two-sublattice configuration represented as $(Ni, Cu)_A(Ti, Zr, Hf)_B$. In this structure, elements occupy specific sublattices rather than acting as substitutional solutes governed by steric constraints. The partitioning behavior is primarily driven by phase equilibria, site preference, the mutual mixing enthalpies of the components on their respective sublattice, and by the relative chemical affinities among the constituents, rather than by atomic mismatch or by the melting points of the pure elements. Cu, which should substitute for Ni on the A-sublattice, exhibits a low partition coefficient ($k \ll 1$) during the initial solidification stage, leading to its strong rejection into the remaining liquid and severe interdendritic segregation. Conversely, Ni is preferentially incorporated into the primary dendritic cores ($k > 1$). Microstructural characterization revealed that Hf segregates into the dendritic cores alongside Ni, despite its structural placement on the B-sublattice with Ti and Zr. This highlights the dominant role of chemical affinity over simple sublattice stoichiometry. The strongly negative mixing enthalpy of Ni–Hf pairs ($\Delta H_{mix}$ = ~ -56 kJ/mol, $x_{Hf}$ = 0.5 in the liquid state) provides a high thermodynamic driving force for their joint solidification in the primary dendrites, exceeding that of Cu-containing pairs. In contrast, Zr is enriched in the interdendritic regions, which follows Cu as the second lowest in the partitioning sequence. Consequently, the effective hierarchy governing elemental partitioning follows the sequence of: Cu < Zr < Ti < Hf < Ni in terms of $k$ for studied alloy in our experiment. Because the partition coefficients of Ti and Hf change sign in their deviation from unity during solidification (**Fig. 4 b**), this ranking should be understood as applying to the initial stage only. To theoretically validate these solidification mechanisms and trace the exact chemical evolution of the residual melt, non-equilibrium Scheil-Gulliver simulations were utilized for all components (**Fig. 4 b**). The model confirms that as solidification progresses, the concentration of Cu in the liquid phase increases sharply due to its extremely low partition coefficient with respect to the primary BCC_B2 phase (**Fig. 4 b**). At the final stage (approximately 860 °C), the residual melt reaches over 53 at. % of Cu, confirming its massive rejection from the dendritic lattice. In contrast, Ni acts as a primary lattice-forming element ($k > 1$), depleting rapidly from the melt to form Ni-rich dendritic cores (**Fig. 4 b**). The solid BCC_B2 phase preferentially consumes Ni, leading to its rapid depletion in the melt and resulting in Ni-rich dendritic cores. The solidification path of Ti displays a unique inflection (**Fig. 4 b**). Initially (down to approximately 1250 °C), the Ti concentration in the melt slightly decreases toward 21 at. %, indicating its early participation in the BCC lattice construction. Upon lattice saturation and further cooling, the trend reverses, and Ti is continuously rejected back into the liquid, accumulating to over 25 at. % at 1000 °C. This dual behavior explains why Ti is partially incorporated within the dendrites but predominantly remains in the interdendritic regions (**Fig. 1 a**). For Zr, the liquid concentration drops steeply to near-zero by 1200 °C (**Fig. 4 b**), which indicates strong partitioning of Zr into the primary BCC_B2 phase. However, our experimental SEM-EDS microstructural analysis reveals that Zr is predominantly segregated into the interdendritic regions. This discrepancy highlights a classic limitation of the thermodynamic model, which overestimates the chemical affinity of Zr in the BCC_B2 phase while underestimating the severe elastic lattice strain induced by the large atomic radius of Zr. Consequently, steric strain effects dominate during real non-equilibrium solidification, forcing Zr out of the dendritic cores and concentrating it into the interdendritic liquid. Lastly, Hf exhibits a transition (**Fig. 4 b**). It initially remains partitioned within the liquid, but below 1200 °C, it massively enters the solid BCC_B2 phase. This explains its presence in the dendrites. At final stage of crystallization, Scheil-Gulliver simulations predict precipitating Hf as the Cu51Hf14 phase. While the Scheil model successfully clarifies the initial microstructural state of the as-cast alloy, understanding the structural stability and the localized melting phenomena during subsequent solid-state compaction requires a transition from solidification history to temperature-dependent phase stability. Therefore, to evaluate how these elemental constituents and the liquid fraction dynamically evolve under near-equilibrium thermal conditions, the equilibrium maps must be systematically analyzed.

Although SEM-BSE observations indicate that the atomized powder particles are macroscopically homogeneous, the application of the non-equilibrium Scheil–Gulliver model remains justified and provides key thermodynamic insights into the system's intrinsic partitioning behavior. Rather than representing a static initial state of the rapidly solidified powder, the Scheil model defines the thermodynamic driving forces and the energetically favorable partitioning pathways along which the alloy tends to evolve. Rapid atomization kinetically freezes the structure into a metastable supersaturated solid solution, suppressing long-range diffusion. However, sub-micron chemical fluctuations and localized excess energy at particle contact zones and grain boundaries remain. During SPS heating up to 1100 °C, thermally activated short-range diffusion, which was enhanced by high defect densities and boundary/pipe diffusion mechanisms, drives solute transport along these thermodynamically preferred paths. Consequently, Cu remains preferentially retained within the particle cores, whereas Ti partitions toward the interparticle networks. This localized redistribution creates micro-regions enriched in Cu (even exceeding 41 at. % in some areas). The local solidus temperature in these Cu-rich zones is significantly decreased, which initiates localized melting at 1100 °C.

The structural evolution can therefore be described as follows. The alloy is first melted at a high heating rate, atomized, and subsequently consolidated in the solid state by SPS. The non-equilibrium solidification path predicted by the Scheil–Gulliver model reveals a strong tendency for chemical partitioning, with progressive rejection of solute elements into the remaining liquid during solidification. Although rapid atomization suppresses long-range diffusion and preserves the powder particles in a metastable, macroscopically homogeneous state, localized solute-enriched regions remain along sub-grain boundaries and particle surfaces. During SPS heating to 1100 °C, the thermodynamic driving force for eliminating supersaturation and reducing interfacial energy activates short-range diffusion in the solid state. Despite the high heating rate, diffusion is kinetically feasible because particle necks and grain boundaries contain high defect densities and excess interfacial energy, promoting grain-boundary and pipe diffusion, which are substantially faster than lattice diffusion. In addition, the high current density and localized electric field characteristic of SPS further facilitate solute transport [48]. As a result, Cu remains preferentially concentrated within particle interiors, whereas Ti partitions toward the interparticle regions. This redistribution produces chemically heterogeneous regions with locally different solidus temperatures. In Cu-enriched regions, where the Cu content increases to approximately 25 – 41 at. % relative to the nominal composition of $Cu_{15}Ni_{35}Ti_{25}Zr_{12.5}Hf_{12.5}$ (at. %), the local solidus temperature is sufficiently reduced to initiate localized melting at 1100 °C. The composition of these low-solidus regions is expected to shift towards the composition of the final residual liquid predicted by the Scheil–Gulliver simulation during non-equilibrium solidification. Therefore, the evolution of both the residual liquid composition and the liquid fraction within the high-temperature range of the Scheil simulation (**Figs. 4 a, b**) provides a quantitative basis for identifying the phases most susceptible to localized melting during SPS. Local overheating at particle contacts caused by the pulsed current cannot be excluded on the basis of post-processing characterization alone, since both mechanisms would produce a resolidified dendritic microstructure. Two observations favor solute-induced decreasing of the local solidus. First, melting is spatially associated with regions of measured Cu enrichment rather than with contact necks in general, whereas discharge-induced heating would be expected at all necks irrespective of local composition. Second, the compositions of the melted regions align with the liquid-forming regime identified independently in **Fig. 5 a** (see below). A definitive differentiation between the two mechanisms would require a reference cycle with the same thermal history but without pulsed current, which was outside the scope of the present work. Preliminary post-SPS annealing experiments performed in a conventional furnace, i.e., without pulsed current, indicate that local melting can also occur at 950 °C after prolonged annealing (48 h), whereas no melting was observed after 24 h. This observation supports the chemically driven, time-dependent nature of the local melting and will be addressed in detail in a forthcoming publication.

**Temperature-Dependent Elemental Redistribution and Liquid Phase Formation**

Thermodynamic calculations were performed over a temperature range of 900 – 1100 °C and an elemental concentration range of 0 to 40 mol. % (**Figs. 5 a – e**). The heat maps were calculated as grid equilibria. The thermodynamic equilibrium at each node was determined by minimizing the total Gibbs energy ( $G_{sys}$):

$$G_{sys} = \sum_{\alpha} n^{\alpha} G_m^{\alpha} \rightarrow min$$

where $n^{\alpha}$ represents the molar amount of phase $\alpha$ and $G_m^{\alpha}$ is the molar Gibbs energy of the respective phase. Equal chemical potentials of all components ($\mu_i^{\alpha} = \mu_i^{LIQUID}$) across coexisting phases were enforced to guarantee phase equilibrium. To model non-ideal interactions in the complex concentrated system (Cu-Ni-Ti-Zr-Hf), the excess Gibbs energy ($G_m^{ex}$) of the phases is defined using Redlich-Kister polynomial expansion:

$$G_m^{ex} = \sum_{i} \sum_{j>i} x_i x_j \sum_{v=0}^{n} L_{i,j}^{v} (x_i - x_j)^v$$

where $x_i$ and $x_j$ are the mole fractions of components *i* and *j*, and $L_{i,j}^{v} = A_v + B_v T$ represents the temperature-dependent binary interaction parameter retrieved from the thermodynamic database. The heat maps capture the temperature-dependent evolution of element concentrations and liquid phase fractions. Within the marked region of interest (**Figs. 5 a – e**), it is evident that at 1100 °C, extensive melting occurs in the green-shaded regions, corresponding to a liquid fraction of 0.5. Under these conditions, critical concentration regimes associated with liquid phase formation are identified as > 25 mol % for Cu, low Ni contents (0 – 26 mol. %), 0 – 17 mol. % and 30 – 40 mol. % for Ti, > 18 mol.% for Hf, and < 5 mol. % together with > 22 mol. % for Zr. Based on these heat maps (**Figs. 5 a – e**) and the chemical analysis (**Figs. 2 a – e**), localized melting is primarily attributed to high Cu concentrations (measured up to 25 – 41 mol. %), with Ti (30 mol. %) and Hf (locally exceeding 18 mol. %) acting as contributing eutectic partners. In contrast, because Ni and Zr concentrations remained outside the critical liquid-forming SPS regimes, they acted as a stable matrix and did not contribute to liquid phase formation. The line scan analysis of the sample SPS-processed at 950 °C for 20 min reveals localized compositional shifts. Dark microstructural regions exhibit a sharp increase in Ti content alongside a slight reduction in Cu (**Fig. 5 f**). Evidence of compositional fluctuations is further highlighted in the sample melted during SPS at 1100 °C (**Fig. 5 g**), where dark areas are enriched in Ti, and the internal black regions are heavily dominated by Cu. Furthermore, Hf displays a discontinuous profile, indicating element redistribution in these regions. Although Ni and Zr variations are present, they do not enter the critical concentration ranges required for liquid phase formation. To illustrate the localized melting process at elevated temperature, a representative equilibrium state at 1100 °C with a Cu concentration of e.g. $x_{Cu}$ = 35 mol. % was selected. The total system size was fixed at $N_{total}$ = 1 mol. Incorporating 35 mol. % of Cu while maintaining the original stoichiometric ratios among the remaining elements (Ni:Ti:Zr:Hf = 35:25:12.5:12.5) yields the following elemental molar constraints ($N_i = \sum_{\alpha} n^{\alpha} x_i^{\alpha}$): $N_{Cu}$ = 0.35 mol, $N_{Ni}$ = 0.2676 mol, $N_{Ti}$ = 0.1912 mol, $N_{Zr}$ = 0.0956 mol, and $N_{Hf}$ = 0.0956 mol. The thermodynamic solver minimizes the total Gibbs energy by evaluating the excess Gibbs energy via the Redlich-Kister expansion for the multicomponent liquid and solid phases. At 1100 °C and 35 mol. % Cu, the system enters a three-phase region (liquid, BCC B2 and a C15b-type Ni5Zr phase), yielding the equilibrium phase amounts: $n^{LIQUID}$ = 0.33 mol and $\sum n^{solid}$ = 0.67 mol (BCC B2 = 0.60 mol, Ni5Zr = 0.07 mol). The coexisting solid and liquid phases satisfy the fundamental phase equilibrium criteria, with identical chemical potentials for each component across the phase interface. These calculations demonstrate that 35 mol. % Cu significantly lowers the solidus temperature of the alloy. At 1100 °C, approximately one third (∼33 %) of the system converts into a liquid phase strongly enriched in Cu (∼73 at. %) and Hf (∼17 at. %), confirming that high Cu concentrations promote extensive melting due to strong thermodynamic interactions in the liquid state, as described by the Redlich-Kister parameter. The localized melting observed at 1100 °C was not caused by the elevated temperature alone, but rather by local chemical inhomogeneity. Elevated local Cu concentrations (above ∼25 mol. %) reduced the melting temperature of these specific regions below 1100 °C. This behavior indicates the formation of low-melting phases or eutectic mixtures, as both Cu–Ti and Cu–Hf systems typically exhibit notably lower eutectic melting points [49-51]. A comparison of the calorimetric data for the Cu–Ti and Cu–Hf systems at 1100 °C [52] confirms the higher exothermicity of the Cu–Hf system. During the dissolution of the solid metal into the melt ($\Delta H_{mix}^{s-l}$), the measured enthalpies for Cu–Ti reach only about -1.25 kJ/mol, whereas for Cu-Hf they decrease to -8.1 kJ/mol. When evaluated relative to the liquid components ($\Delta H_{mix}^{l-l}$), the minimum for Cu–Ti is approximately -4.2 kJ/mol, compared to -19.2 kJ/mol for Cu-Hf. Both cases clearly demonstrate a significantly stronger exothermic interaction in the Cu–Hf system. The composition of the potential melt is shown in **Fig. 5 h**. The figure clearly demonstrates that thermodynamic modeling confirms that Cu is the primary driver of the localized melting, partitioning strongly into the liquid phase (∼ 38 at. % Cu at 1100 °C compared to 15 at. % nominally). Ti exhibits near-neutral partitioning, maintaining a liquid concentration near its nominal level (∼ 25 at. %), while Hf is moderately enriched (∼ 23 at. %). Conversely, Ni and Zr remain predominantly in the solid matrix (**Fig. 5 h**). Hf and Ti therefore do not act as independent triggers of melting. They are the eutectic partners of Cu that are already present in the Cu-rich regions, with the Cu–Hf pair providing the dominant contribution to the solidus depression owing to its considerably more negative mixing enthalpy. The absence of melting at 950 °C, despite Cu concentrations of up to 41 at. % measured after SPS at 950 °C, is therefore a consequence of temperature rather than of a missing alloying partner. The local solidus of the Cu-enriched regions lies above 950 °C but below 1100 °C (calculated local solidus for the measured compositions for the region measured after SPS at 950 °C for 10 min containing 41.6 at. % Cu, 27 at. % Ni, 11.8 at. % Ti, 11.9 at. % Zr, and 7.7 at. % Hf was determined to be 1087 °C). Notably, this region contains less Hf than the nominal alloy, confirming that Cu enrichment alone is sufficient to depress the local solidus below 1100 °C, with Hf and Ti acting as partners present at moderate concentrations. At 950 °C, diffusion was clearly active, as evidenced by Ti and Hf enrichment and Cu depletion along boundaries and dark regions, yet the Cu-rich regions remained below their local solidus. Upon reaching 1100 °C, these regions crossed their local solidus, resulting in the formation of a liquid phase. Melting observed already at 1000 °C indicates that, at this temperature, some regions had reached compositions with an even lower local solidus than the analyzed one, e.g., higher Cu contents combined with Hf or Ti enrichment (**Figs. 5 a, c, d**). Throughout this entire process, Ni and Zr acted as a stable matrix, showing no tendency toward local segregation that would induce melting within this temperature window.

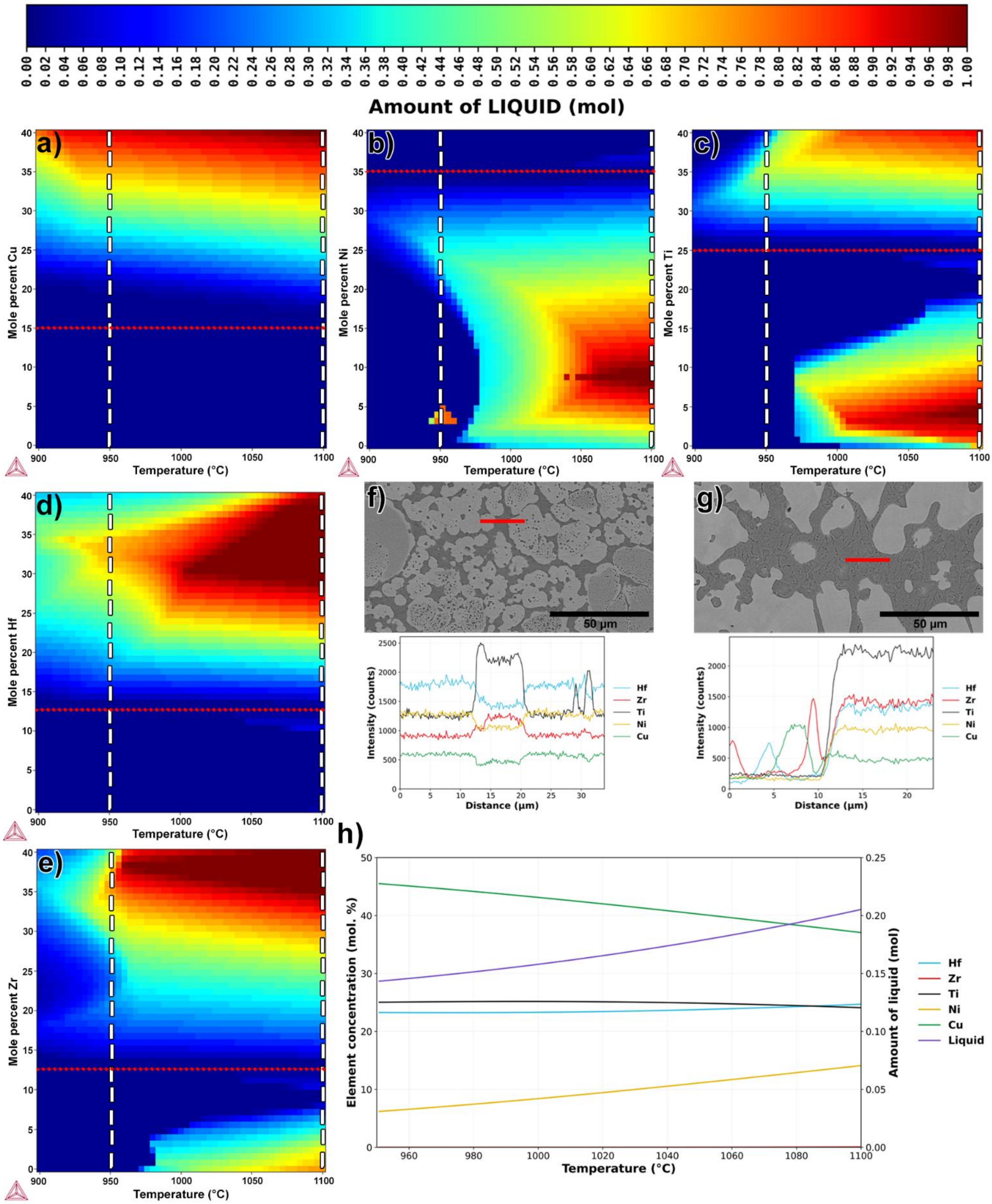

**Fig. 5:** a – e) Heat maps for a) Cu, b) Ni, c) Ti, d) Hf, and e) Zr, where red dashed lines indicate the initial composition and white dashed outlines mark the regions of interest. f, g) microstructure and EDS intensity profile scans of SPSed samples prepared at f) 950 °C for 20 min and g) 1100 °C for 5 min, h) concentration of individual elements and liquid phase fraction as a function of temperature for Cu15Ni35Ti25Zr12.5Hf12.5 (at. %) alloy during non-equilibrium (Scheil–Gulliver) solidification

The premature melting of the powder compacts below the nominal equilibrium solidus can be consistently explained. The as-cast alloy exhibits a strong tendency towards chemical segregation, as predicted by the Scheil model (**Fig. 4 a**). Although subsequent atomization kinetically suppresses this segregation due to ultra-high cooling rates, it produces a metastable supersaturated powder structure that retains a thermodynamic driving force for elemental segregation along the initial solidification pathways. During SPS heating up to 1100 °C, this metastability drives rapid short-range diffusion in the solid state. A key characteristic of this in-situ separation is the preferential migration of Ti into the continuous interparticle boundary network after SPS, driven by the thermodynamic reduction of interfacial free energy. Simultaneously, Cu remains strongly retained and concentrated within the spherical particle cores and adjacent contact zones, reaching localized concentrations up to 41 at. % Cu after SPS. Because this enrichment largely reflects Cu depletion of the surrounding regions, it does not require long-range Cu transport.

The structural effects of this pronounced chemical partitioning are quantified by the 2D local equilibrium calculations (**Fig. 5 a**). While the nominal alloy composition is predicted to be fully solid at 1000 °C and to contain only ∼0.1 % of liquid at 1100 °C (**Fig. 3 b**), Cu enrichment above ∼25 at. % markedly lowers the local solidus. Upon reaching 1100 °C, these Cu-enriched regions exceed their local solidus, which is depressed by the strongly negative mixing enthalpy of the Cu–Hf pair, resulting in localized melting. These results indicate that thermally activated solid-state partitioning, rather than macroscopic thermal gradients, is the primary factor controlling premature melting during SPS. Furthermore, these localized contact zones, characterized by elevated interfacial energy and intense local solute accumulation, represent highly favorable kinetic sites that accelerate liquid formation within these chemically susceptible micro-regions.

The elemental distribution maps reveal a distinct spatial separation between Cu and Hf during solid-state sintering at temperatures up to 950 °C (**Fig. 2 b – d**). In the solid state, solute partitioning is governed by boundary energy minimization and strain-driven kinetics, concentrating Ti and Hf along interparticle networks while Cu remains preferentially localized within particle interiors after SPS. Cu–Hf co-segregation is therefore not a prerequisite for melting but its consequence. Once the local solidus is crossed at 1100 °C, solid-state lattice constraints are removed, and the strong chemical affinity between Cu and Hf in the liquid drives their co-segregation during resolidification, as reflected by the dendrites formed at 1100 °C containing Cu together with 21.1 at. % Hf (**Fig. 2 e**). The composition of the last liquid to solidify is expected to approach that of the terminal reaction predicted by the Scheil–Gulliver simulation, in which the Cu51(Hf, Zr)14 phase forms. The presence of this phase in the SPS-processed material was not verified in the present work and its identification remains a subject for further study. The 2D local equilibrium map for Hf (**Fig. 5 d**) further shows that local Hf enrichment, as detected by EDS after SPS, additionally promotes liquid formation at 1100 °C. However, as demonstrated by the local solidus calculation, it is not a prerequisite for melting, and Cu remains its primary trigger.

For the present alloy, SPS should therefore not exceed 950 °C, and subsequent annealing at this temperature should be limited to max. 24 h (**Fig. 6**). More generally, the safe processing window of rapidly solidified CCA powders containing a low-melting constituent can be estimated from two thermodynamic bounds: i) the end of Scheil solidification as a conservative lower limit and ii) the solidus of the measured chemical heterogeneities as an upper limit, while the dwell time must be restricted to avoid progressive local enrichment.

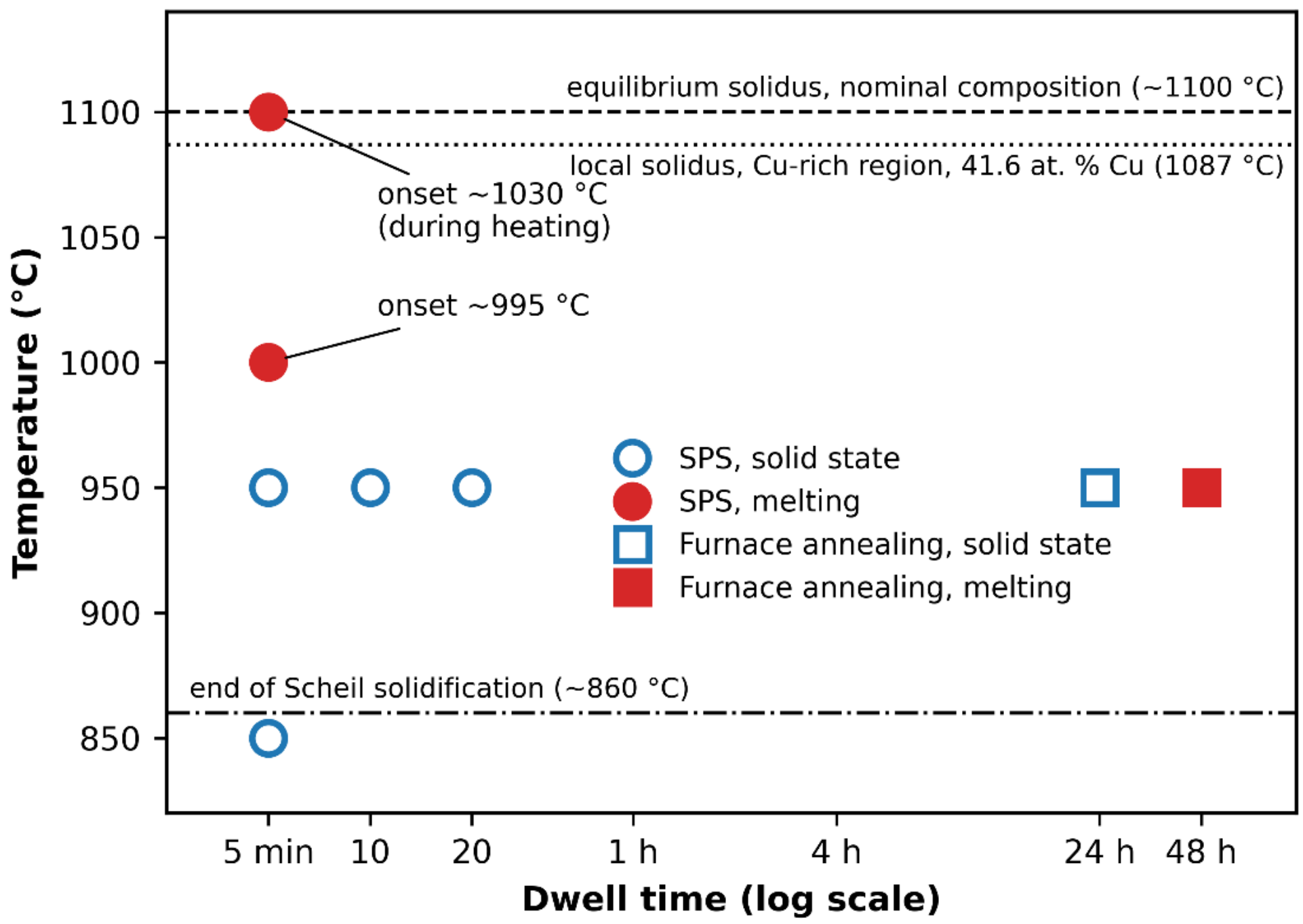

**Fig. 6:** Processing map of the Cu15Ni35Ti25Zr12.5Hf12.5 (at. %) powder: SPS (circles) and preliminary post-SPS furnace annealing (950 °C, 24 h and 48 h, Ar) (squares); open symbols denote solid-state processing, filled symbols melting. Horizontal lines mark the equilibrium solidus of the nominal composition, the local solidus of the Cu-rich region and the end of Scheil solidification

#### Conclusion

Although equilibrium calculations using the current TCHEA7 database predict a fully solid state of the nominal composition up to slightly below 1100 °C (and the earlier TCHEA4 version up to ~1260 °C), localized melting occurred during SPS already at 1000 °C and at 1100 °C. Retrospective Scheil–Gulliver simulations combined with 2D local equilibrium maps revealed that this behavior originates from non-equilibrium microsegregation inherited from powder atomization. During SPS heating, Ti partitions towards the interparticle network, whereas Cu is retained within the particle interiors. Owing to the strongly negative mixing enthalpy of Cu with Hf and, to a lesser extent, with Ti, these Cu-rich regions exhibit a local solidus well below the nominal value; upon reaching 1100 °C, they cross it and form a low-melting Cu-based liquid. The Cu–Hf co-segregation observed after melting is a consequence of the solidification of this liquid rather than its cause. These findings show that thermally activated solid-state partitioning of a low-melting element with a strongly negative mixing enthalpy towards other principal constituents, rather than macroscopic thermal conditions, determines the processing limits of complex concentrated alloy powders.

#### Declaration of Interest Statement

The authors declare that they have no known competing financial interests or personal relationships that could have appeared to influence the work reported in this paper.

#### Acknowledgment

The authors acknowledge financial support from the Czech Science Foundation (Grant No. 26-21374S). The authors acknowledge the assistance provided by the Ferroic Multifunctionalities project, supported by the Ministry of Education, Youth, and Sports of the Czech Republic. Project No. CZ.02.01.01/00/22_008/0004591, co-funded by the European Union. Measurements of chemical composition were performed using the CzechNanoLab research infrastructure, supported by MEYS CR (LM2023051).

**Data availability**

The used data are accessible via the Zenodo repository: https://doi.org/10.5281/zenodo.23398114

**Declaration of generative AI and AI-assisted technologies in the manuscript preparation process**

During the preparation of this work the authors used Gemini by Google, ChatGPT by OpenAI or Claude by Anthropic in order to assist with language polishing, grammar correction, and improvement of textual clarity and flow. After using this tool, the authors reviewed and edited the content as needed and take full responsibility for the content of the published article.